\documentclass[12pt]{elsarticle}

\usepackage{amssymb}
\journal{Ultramicroscopy}
\usepackage{hyperref}
\usepackage{xurl} 
\usepackage{xcolor}
\usepackage{algorithm}
\usepackage{algpseudocode}
\usepackage{textcomp}
\usepackage{amsmath}
\usepackage[table]{xcolor}
\makeatletter
\def\ps@pprintTitle{%
  \let\@oddhead\@empty
  \let\@evenhead\@empty
  \let\@oddfoot\@empty
  \let\@evenfoot\@oddfoot}
\makeatother
\begin{document}

\begin{frontmatter}



\title{An Interactive, Automated 4D-STEM data acquisition and analysis routine for Scanning Electron Nanobeam Diffraction and Ptychography experiments}


\author[inst1]{Mohsen Danaie}

\affiliation[inst1]{organization={electron Physical Science Imaging Centre},%Department and Organization
            addressline={Diamond Light Source}, 
            city={Didcot},
            postcode={OX11 0DE}, 
            state={Oxfordshire},
            country={United Kingdom}}

\author[inst1]{Max England}
\author[inst1,inst3]{Yiming Xu}
\author[inst1]{Ruomu Zhang}
\author[inst3]{Ed Darnbrough}
\author[inst1]{Josh Willem De Boer}
\author[inst1]{Frederick Allars}
\author[inst3,inst5]{Zaeem Najeeb}
\author[inst5]{Aakash Varambhia}
\author[inst1]{Jinseok Ryu}
\author[inst2]{Benjamin Bradnick}
\author[inst4]{Damien McGrouther}
\author[inst5]{Manfred E. Schuster}
\author[inst1]{Christopher S. Allen}

\affiliation[inst2]{organization={Quantum Detectors},%Department and Organization
            addressline={Harwell Campus, R103}, 
            city={Oxford},
            postcode={OX11 0QX}, 
            state={Oxfordshire},
            country={United Kingdom}}
\affiliation[inst3]{organization={Department of Materials},%Department and Organization
            addressline={University of Oxford}, 
            city={Oxford},
            postcode={OX1 3PH}, 
            state={Oxfordshire},
            country={United Kingdom}}

\affiliation[inst4]{organization={JEOL UK},%Department and Organization
            addressline={JEOL House}, 
            city={Welwyn Garden City},
            postcode={AL7 1LT}, 
            state={Herts.},
            country={United Kingdom}}
            
\affiliation[inst5]{organization={Johnson Matthey},%Department and Organization
            addressline={Johnson Matthey Technology Centre}, 
            city={Sonning Common},
            postcode={RG4 9NH}, 
            state={Berkshire},
            country={United Kingdom}}

\begin{abstract}
Modern transmission electron microscopes are versatile instruments which have become indispensable tools for understanding structure and chemical composition at the nano- and atomic scale. In the physical sciences these instruments are still largely manually controlled, requiring significant operator expertise, limiting throughput, and precluding statistical analysis of large datasets. Recent technical advances in both hardware and in control software now allow for the interaction with almost every functionality of the microscope through a programming interface.  This enables better experimental design and data collection automation while also reducing operator collection bias and required expertise.  In this study, we present an automated data collection routine with machine-driven decision-making to enable the collection of hundreds of 4D-STEM nanobeam diffraction and ptychography data from a large distribution of size-selectively deposited Pt nanoparticles. We present a semi-automated data analysis workflow to extract pertinent information from the large volumes of collected data. For the nanobeam diffraction data, reducing each dataset to its azimuthal variance profile and combining automated crystal orientation mapping with per-particle morphology descriptors reveals the orientation, shape and phase distributions across the ensemble, including a weak \{110\} texture. For the ptychographic data, an automated screening pipeline identifies on-zone-axis particles and enables atomic-resolution phase imaging and lattice-strain mapping of individual grains. Together these demonstrate how automation turns instrument throughput into statistically meaningful, atomic-scale microstructural information.
\end{abstract}

\begin{graphicalabstract}
\includegraphics[width=\textwidth]{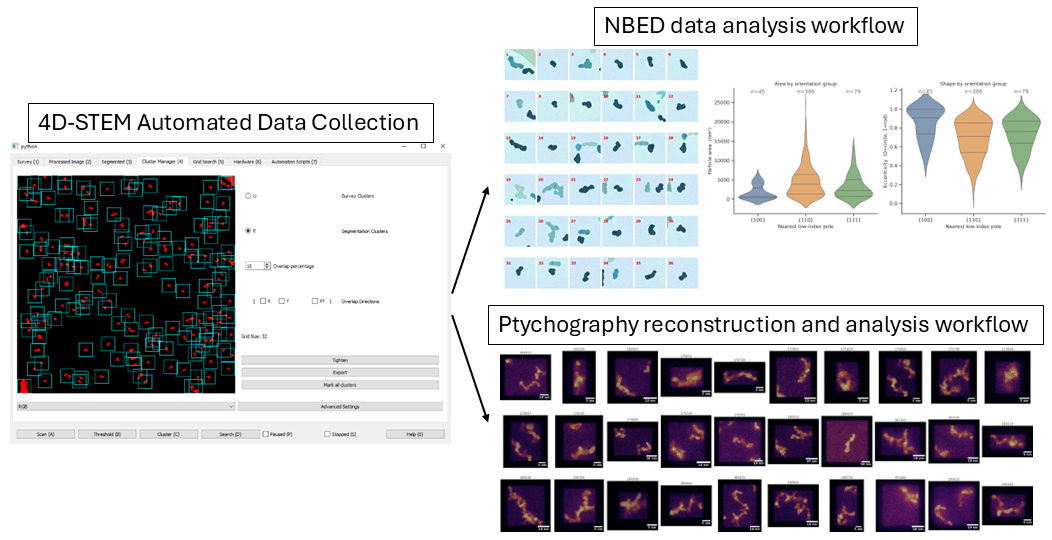}
\end{graphicalabstract}

\begin{highlights}
\item Automating 4D-STEM data collection with user-defined criteria
\item Experimental demonstration of data collection automation both in nanobeam and in ptychography modes of 4D-STEM
\item Data analysis workflow in reducing the data dimensionality to reach rich statistical information on the microstructure
\end{highlights}


\end{frontmatter}


\section{Introduction}
\label{sec:intro}

The transmission electron microscope (TEM) is one of the most versatile instruments available to the physical sciences experimentalist, capable of probing structure, chemistry, and bonding from the micrometre scale down to the level of individual atomic columns. In a typical physical sciences experiment, the operator has to navigate a structurally heterogeneous sample -- nanoparticles on an amorphous support film, a focused-ion-beam prepared lamella, or a polycrystalline thin foil -- and identify regions that are both electron-transparent and representative of the phenomenon under study. Once a suitable region is found, the operator then selects the appropriate imaging or diffraction mode, align the optics, set the illumination conditions, and adjust beam aberrations, all before a single frame of data is recorded. In diffraction-contrast TEM, for example, the precise orientation of the crystal with respect to the electron beam must be controlled manually to bring a specific reflection or zone axis into the desired diffraction condition, a process that demands both instrument operator expertise and considerable time \cite{williams2009}. For scanning TEM (STEM) mode, additional alignment and calibration steps -- probe correction, scan rotation calibration, and detector collection angles -- must be completed before STEM images or spectroscopic signals such as energy-dispersive X-ray spectroscopy (EDX) or electron energy-loss spectroscopy (EELS) can be acquired. Furthermore, over the course of an experimental session the operator needs to frequently correct for low-order aberrations -- 2D astigmatism, coma and defocus -- to maintain data quality. This operator-intensive workflow limits experimental throughput, introduces collection bias, and makes statistically meaningful surveys across large numbers of particles or grain orientations practically inaccessible.

Recent years have seen significant advances in the hardware available on modern TEMs that are beginning to change this picture. Perhaps the most transformative development has been the widespread adoption of fast, direct electron detectors capable of acquiring full two-dimensional diffraction patterns at every probe position in a STEM scan -- so-called four-dimensional STEM (4D-STEM) \cite{4d_stem_review_ophus}. These hybrid-pixel and direct-detection detectors \cite{mcmullan2016direct, clough2016direct} can operate at frame rates of hundreds to thousands of frames per second, generating datasets that would have been inconceivable on the slow charge-coupled-device cameras that preceded them. These hardware advances produce data at rates and volumes that manual operation can no longer keep pace with; a single 4D-STEM session can yield hundreds of datasets within a few hours, far exceeding what an operator could sensibly acquire and review interactively.

Alongside hardware improvements, microscope and detector vendors have begun exposing application programming interfaces (APIs) that allow external software to query and control almost every function of the microscope and its ancillary hardware. Libraries such as \textit{PyJEM} for JEOL instruments \cite{pyjem} provide Python-level access to lens and deflector excitations, stage positions, magnification, and lens parameters. Scan engines with open interfaces \cite{qd_scan_engine} allow arbitrary user-defined scan patterns to be executed, decoupling the data acquisition trajectory from the constraints of the native microscope scan controller. Detector manufacturers similarly provide TCP/IP command sets for remote configuration of thresholds, bit depth, and frame triggering. Together, these interfaces create the software foundation on which fully automated, closed-loop experiments can be built.

In the biological sciences, the potential of such automation was recognised much earlier and has been exploited for decades. Single-particle cryo-electron microscopy (cryo-EM) relies fundamentally on automated data collection: the sample consists of many thousands of identical copies of the same protein complex, frozen in random orientations in a thin vitrified ice film, and the goal is to record enough micrographs from enough different regions to reconstruct a three-dimensional density map by averaging \cite{Vinothkumar_Henderson_2016, ChengYifan2018Scdi}. The sample homogeneity and the statistical nature of the reconstruction problem make automation both tractable and essential -- no operator could manually acquire the tens of thousands of images needed for a high-resolution structure in a reasonable time. Sophisticated software packages have been developed to drive the microscope autonomously through the entire data collection session, performing autofocus, beam-tilt correction, hole-finding, and image acquisition without human intervention. The result is that cryo-EM facilities routinely run unattended overnight, and the technique has become a high-throughput structural biology platform \cite{ChengYifan2018Scdi}.

Physical sciences EM lags significantly behind this standard. The heterogeneity of physical science samples, the diversity of operation modes, and the need for expert judgement at multiple stages of the experiment, plus the need to frequently adjust for lower order aberrations (2D-astigmatism, coma and defocus), have historically made automation more challenging. An additional layer of complexity arises in \textit{operando} and \textit{in-situ} experiments -- heating, gas-cell, or electrochemical-cell holders -- where environmental conditions must also be controlled in concert with the microscope. Despite these challenges, the case for automation is compelling: the hardware advances described above mean that the bottleneck in physical sciences EM is increasingly the operator rather than the instrument. Without automated data collection the statistical power latent in fast pixellated detectors cannot be realised, since the number of datasets an operator can manually set up and acquire in a session is far smaller than the instrument can in principle deliver. Recent work has begun to address this gap, with demonstrations of machine-learning-guided closed-loop STEM \cite{spurgeon_autoSTEM}, variational-autoencoder-driven automated atomic-resolution imaging \cite{Creange_2022, Ziatdinov2023}, and broader perspectives on data-driven next-generation electron microscopy \cite{Spurgeon2021}.

In the present work, we describe a modular, open-source automated data collection framework built on the hardware APIs available on a JEOL GrandARM 300F at the electron Physical Sciences Imaging Centre (ePSIC), Diamond Light Source. The software incorporates a graphical user interface (GUI) so that operators can define collection criteria and monitor progress, and a decision-making algorithm that identifies suitable sample regions, populates them with acquisition tasks, and executes those tasks autonomously. We demonstrate the framework in two complementary modes: scanning nanobeam electron diffraction (NBED), where the goal is statistical characterisation of crystal orientations and morphologies across a large ensemble of nanoparticles; and ptychographic 4D-STEM, where the goal is atomic-resolution phase imaging of selected particles. The sample is an ensemble of size-selectively deposited Pt nanoparticles on an amorphous carbon support -- a specimen comprising hundreds of well-separated particles whose size, shape, and orientation distributions can only be meaningfully characterised through exactly the kind of high-throughput, statistically rich data collection that automation enables.

\section{Methods}
\label{sec:methods}

\subsection{Hardware details}
\label{sec:methods_hw}
\begin{figure} [h]
    \centering
    \includegraphics[width=\columnwidth]{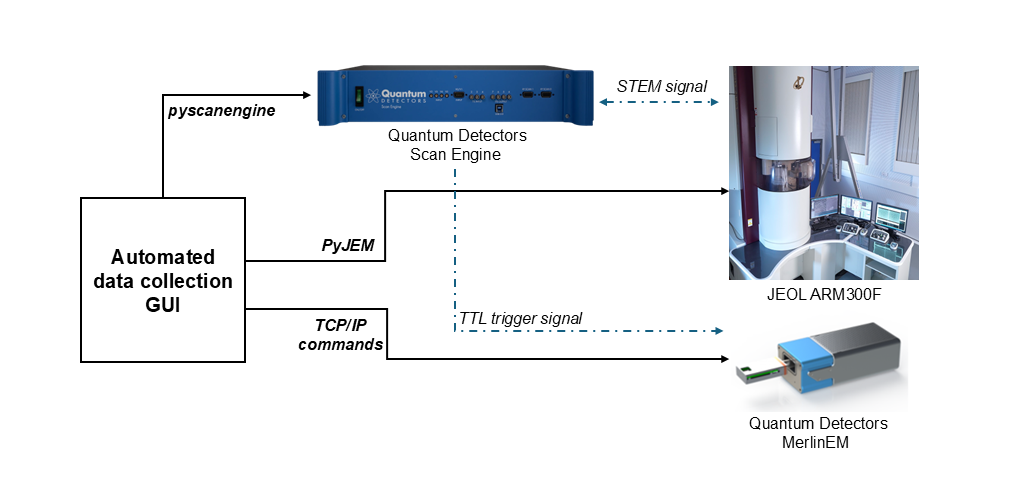}
    \caption{Hardware components and their interactions to enable automated data acquisition. The automated data collection GUI communicates with three main hardware components: the JEOL ARM300F microscope (via \textit{PyJEM}), the Quantum Detectors Scan Engine (via \textit{pyscanengine}), and the Quantum Detectors MerlinEM pixellated detector (via TCP/IP commands). Solid lines represent software communication with each component. The scan engine receives the STEM signal from the microscope and delivers a TTL trigger signal to the MerlinEM detector to synchronise frame capture with the scan.}
    \label{fig:fig_hw}
\end{figure}

Figure~\ref{fig:fig_hw} shows a schematic of the hardware set-up on the microscope and how the automated data collection software communicates with the various components. The control software runs on a support PC on the microscope network and interfaces with three main hardware components: the microscope column, the scan engine, and the pixellated detector.

The microscope, in this case a JEOL GrandARM 300F, is connected via a TCP/IP port and remotely controlled using the \textit{PyJEM} Python library \cite{pyjem}. Through this library the state of almost all functions of the microscope -- lens and deflector excitations, aperture selection, stage position, magnification, and detector insertion -- can be queried and altered as needed. A Quantum Detectors Scan Engine is used to drive the microscope scan coils, and reads the analogue signals of the microscope STEM detectors (in this work the user has a choice between the bright-field and high-angle annular dark-field signals). The scan engine is controlled by a Python package, \textit{pyscanengine} \cite{qd_scan_engine}, enabling the user to create arbitrary scan patterns. Though in this first application only raster scans are used, non-raster patterns can be simply implemented. The scan engine enables arbitrary setting of the scan positions and dwell times as well as enabling larger scan arrays to be performed (max. 16384 sampling positions per axis). Of particular importance to the automated workflow presented here, the scan engine allows an arbitrary rectangular sub-region of the full scan field to be addressed directly, so that successive 4D-STEM acquisitions can be placed anywhere within the survey field of view without changing magnification or moving the stage. The scan engine triggers the pixellated detector via a TTL (transistor-transistor logic) output line to ensure frame-accurate synchronisation of the scan with the detector read-out. A Quantum Detectors MerlinEM quad-chip direct electron detector (four Medipix3 chips tiled in a $2\times2$ configuration, giving an effective $515 \times 515$ pixel array) is used to record the 4D-STEM data. Communication with the detector is through a TCP/IP port. All the detector settings, including counter depth, frame start / stop triggers and noise threshold levels, are set by remote TCP/IP commands, and the acquired frames are streamed to disk on the storage server at the Diamond Light Source.

\subsection{Software implementation}
\label{sec:methods_sw}
\begin{figure}
    \centering
    \includegraphics[width=0.85\columnwidth]{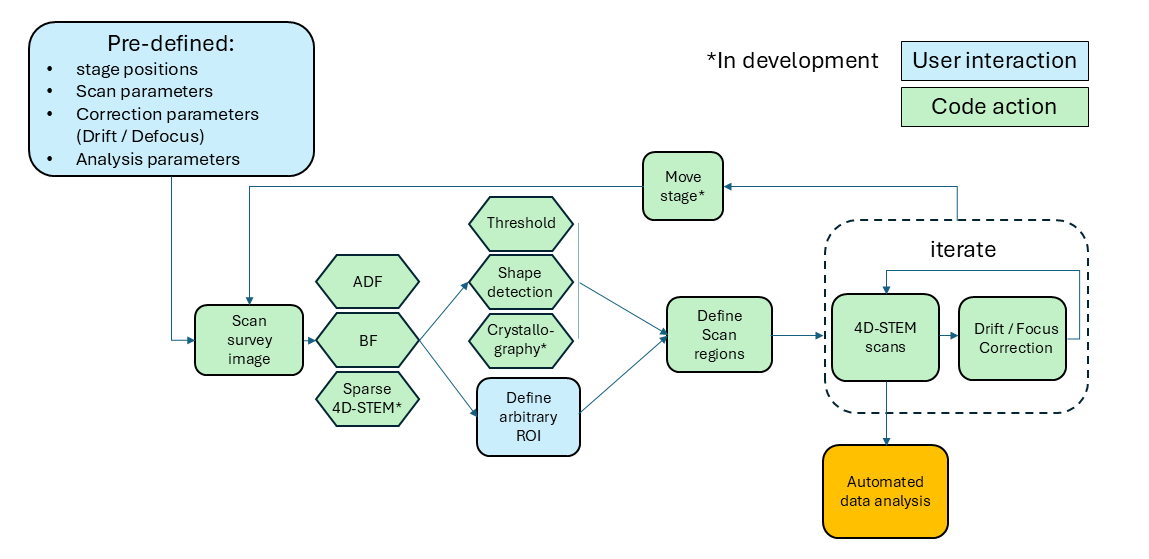}
    \caption{Flowchart of the automated data collection software workflow. Pre-defined parameters (stage positions, scan and correction parameters, analysis settings) are set by the user prior to acquisition. A survey image is acquired using ADF or BF STEM (sparse 4D-STEM is also in development), and analysed via thresholding, shape detection, or user-defined ROI selection to define the 4D-STEM scan regions. An optional stage movement step (in development) enables stitching across larger areas. Within the acquisition loop, 4D-STEM scans are interleaved with drift and focus corrections. The collected data feed directly into an automated analysis pipeline.}
    \label{fig:fig_workflow}
\end{figure}

\begin{figure}
    \centering
    \includegraphics[width=\columnwidth]{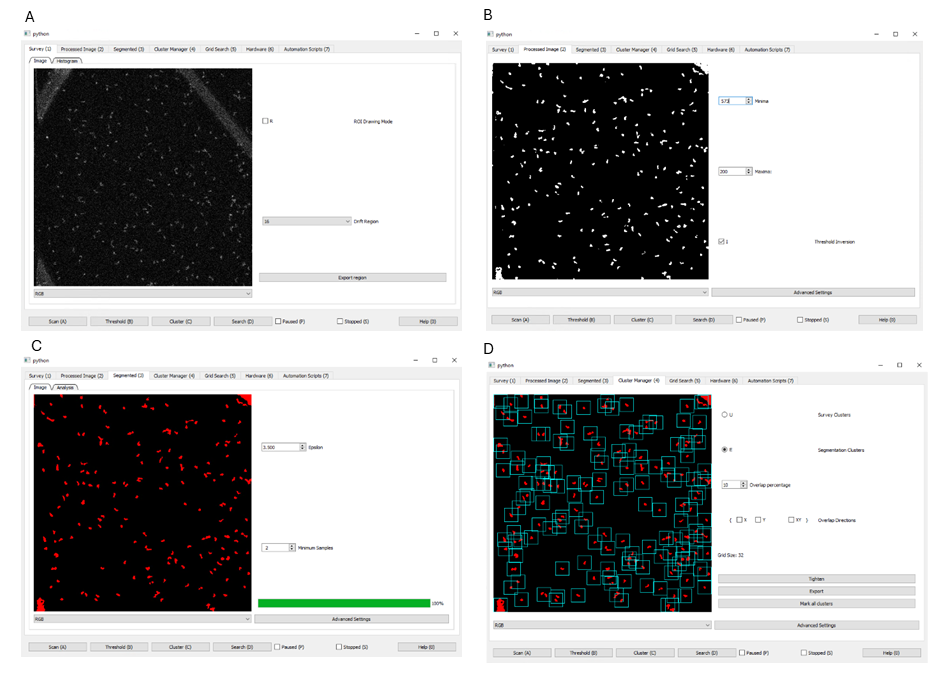}
    \caption{Screenshots of the GUI at successive stages of automated data collection for a Pt nanoparticle sample: \textbf{A} survey HAADF-STEM image, \textbf{B} binarised image after intensity thresholding to isolate nanoparticles, \textbf{C} DBSCAN-clustered binary image with individual particle regions highlighted in red, \textbf{D} 4D-STEM acquisition grid populated over the segmented particle regions with user-defined overlap.}
    \label{fig:fig_sw}
\end{figure}

The Automated Data Collection software with a graphical user interface (GUI), available through the GitHub repository \cite{epsic}, was written in Python using the above packages (\textit{PyJEM}, \textit{pyscanengine} and the remote commands for MerlinEM controls), together with other standard libraries: \textit{PyQt5} \cite{pyqt5} for the interactive GUI front end, \textit{OpenCV} \cite{opencv_library} for image processing, and \textit{scikit-learn} \cite{scikit-learn} for clustering the binarised survey image. The hardware communication is wrapped in dedicated modules behind a common interface, so that the decision-making and GUI layers are agnostic to the specific microscope, scan engine, or detector; porting the software to a different instrument requires only re-implementing these hardware modules. The pseudo-code in \autoref{alg:alg1} shows the step-by-step operations of the software, and the corresponding flowchart is given in Figure~\ref{fig:fig_workflow}. Snapshots in Figure~\ref{fig:fig_sw} show the GUI at various stages of setting up and data collection.

The GUI is organised as a series of tabs that mirror the acquisition sequence -- survey, image processing, segmentation, cluster management, grid search, and hardware settings -- with the most commonly used parameters exposed directly and more specialised options collected behind advanced-settings panels or expressed through a \textit{config} file. A typical data acquisition session proceeds as follows: A survey image is first acquired through the scan engine using one of the analogue STEM signals (annular dark-field in the present work; bright-field is equally applicable). The user then binarises this image by setting lower and upper intensity thresholds against a live histogram view (Figure~\ref{fig:fig_sw}\textbf{B}); for the size-selectively deposited Pt nanoparticle sample used in this work, the particles appear as bright, well-separated features on the amorphous carbon support, making intensity thresholding a robust segmentation criterion. The foreground pixels of the binary image are then grouped into individual particles using the density-based DBSCAN clustering algorithm \cite{dbscan} as implemented in \textit{scikit-learn}, with the neighbourhood radius ($\varepsilon$) and minimum-samples parameters adjustable in the GUI (Figure~\ref{fig:fig_sw}\textbf{C}). Each resulting cluster is then covered by one or more square 4D-STEM acquisition regions of a user-defined size, with a user-defined overlap percentage between neighbouring regions ensuring that particles larger than a single scan field are fully captured (Figure~\ref{fig:fig_sw}\textbf{D}). The populated regions are placed in an acquisition queue for autonomous execution.

During the acquisition loop, the software steps through the queue, re-configuring the scan engine sub-region for each entry, arming the MerlinEM detector, and executing the 4D-STEM scan. At a user-defined frequency, the loop is interleaved with image-based drift measurement and defocus correction steps; the measured drift history is recorded and used to update subsequent scan positions. Acquisition can be paused, resumed, or aborted from the GUI at any point without loss of the already-collected data. Alongside each dataset, the software captures a comprehensive metadata record -- microscope state, scan and detector parameters, stage coordinates, and the survey and thresholded images on which the region-selection decisions were based (see \ref{sec:appendix_metadata}) -- and organises the output into a structured \textit{HDF5} file format hierarchy that feeds directly into the automated analysis pipeline described in Section~\ref{sec:analysis}. For unattended multi-region experiments, a lightweight scripting interface additionally allows sequences of GUI and microscope operations to be composed and replayed without manual interaction.
\begin{algorithm}
\caption{Automated Data Collection Workflow}
\label{alg:alg1}
\begin{algorithmic}[1]
\small
\State Initialize system
    \State Load GUI framework
    \State Connect to microscope using \textit{PyJEM} and verify hardware connections
\State User sets survey and acquisition sampling parameters via GUI
    \State Define data collection regions and imaging parameters
    \State Preview sample with survey image
\State Set up microscope live corrections
    \State Set frequency of focus and drift corrections
\State Collect survey image
    \State Binarise the image by applying a threshold
    \State Cluster the binary image into separate labels
    \State Populate 4D-STEM acquisitions to cover the labels with some user-defined overlap
    \State Queue tasks for execution
\While{regions remain in queue}
    \State Change scan sub-region to the next region
    \If{Time to perform correction}
        \State Perform drift / defocus correction
    \EndIf
    \State Acquire 4D-STEM data
    \State Save data and log metadata
    \If{errors detected}
        \State Pause or adjust workflow
    \EndIf
\EndWhile
\State Perform post-processing
\State Organize data into structured folders
\State Shutdown and clean-up
\end{algorithmic}
\end{algorithm}

\section{Proof of principle experiment}
\label{sec:methods_exp}

\subsection{Materials and sample preparation}
\label{sec:methods_exp_sample}
The test specimen consisted of Pt nanoparticles size-selectively deposited directly onto an amorphous carbon TEM support film. Size-selective deposition was carried out on a dual magnetron cluster source based on the design described in references \cite{von1999new, pratontep2005size}, producing ligand-free particles with a controlled size distribution dispersed over the support at a controlled coverage, yielding a large population of well-separated, nominally similar particles. This makes the specimen an ideal benchmark for automated collection: the individual particles are readily segmented from the support in a survey image, and the scientific questions of interest -- the true distributions of particle size, shape, and crystallographic orientation across the ensemble -- are inherently statistical and can only be meaningfully addressed by sampling hundreds of particles. 

\subsection{Data collection}
\label{sec:methods_exp_data}
For this example experiment the microscope was aligned at 300~kV accelerating voltage in nano-beam electron diffraction (NBED) mode with the probe convergence semi-angle around 1 mrad. To achieve this, the probe corrector optics on the microscope were turned off and the excitations of the probe forming lenses were altered to minimise the probe convergence semi-angle. Probe tilt and shift purity were corrected and scan pattern was adjusted with respect to a reference square pattern in the cross-grating standard specimen.

The sample was the size-selectively deposited Pt nanoparticle specimen described in Section~\ref{sec:methods_exp_sample}.

Prior to the experimental data collection, reference data from standard cross-grating replica sample were collected at the same magnification as the survey image (150~kX). Details are discussed in appendix here ~\ref{sec:appendix_cals}. The screenshots shown in Figure~\ref{fig:fig_sw} depict the region used for automated data collection. The field-of-view at the survey image scan is 1.33~\textmu m and the FOV of the individual 4D-STEM datasets was set to 1/16th of the survey FOV, i.e. 83.3 nm. The nominal camera-length used for the diffraction patterns was 40 cm. Using the standard evaporated gold sample the calibrated reciprocal space pixel size was measured to be 0.005654~\text{\AA$^{-1}$}. An affine transformation correction matrix was also measured using the gold diffraction rings, which was applied to the experimental data. For the survey image the high-angle annular dark-field (HAADF) STEM signal was used with $512 \times 512$ pixel array size in the raster image. Each of the 4D-STEM data collected had the dimensions of $256 \times 256$ in real space and $515 \times 515$ in reciprocal space, collected at the same magnification, i.e. 150~kX, but with the  sampling rate of the QDScan Engine set to 4096, resulting in the real space step size of 0.3255~nm (FOV for each 4D-STEM data was 83.3~nm). Each diffraction pattern was collected on MerlinEM detector with 1~ms dwell time and 6~bit read-out bit depth.

The data collection ran autonomously, with a total of 153 datasets collected. The binarisation and clustering steps of the automated routine identified the individual Pt nanoparticles against the amorphous carbon background, so that 4D-STEM acquisitions were placed only on particle-containing regions of the support and minimal beam time was spent on empty carbon. Given that in the optical configuration used (NBED mode, 1~mrad convergence angle) the depth of focus is estimated to be around 1.5~\textmu m (based on equation (1) in \cite{NELLIST201718}), automated focus correction was not required for data collection in this proof-of-principle experiment. Drift correction was carried out after every 4 acquisitions, using a reference region and a cross-correlation algorithm to measure the drift vectors, shifting the acquisition windows accordingly.

\section{Data analysis workflow}
\label{sec:analysis}
Given the large number of datasets collected, a data analysis workflow was designed around the open-source packages \textit{Py4DSTEM}~\cite{py4dstem_paper}, \textit{hyperspy}~\cite{hyperspy_paper}, \textit{orix}~\cite{orix}, \textit{scikit-image}~\cite{scikit-image} and \textit{scikit-learn}~\cite{scikit-learn}, and executed on the high-performance computing (HPC) cluster at Diamond Light Source. To process every acquisition identically and reproducibly, the per-dataset analysis is written as a single parameterised template notebook; a helper routine then clones this template for each acquisition in the run -- substituting only the dataset path and experimental parameters -- and submits the resulting notebooks to the cluster as a SLURM job array, with a user-defined number of jobs running concurrently. The complete set of analysis notebooks is available in the accompanying repository \cite{epsic_analysis}. For the NBED data, the per-dataset workflow can be summarised as follows:
\\
\textit{a. Prepare raw data}: import the raw MerlinEM data (.mib file format), mask the detector hot pixels and save a version readable by \textit{Py4DSTEM}.
\\
\textit{b. Load data and apply calibrations}: the reciprocal-space pixel size and the elliptical-distortion parameters, measured from the evaporated-gold cross-grating standard (\ref{sec:appendix_cals}), are read from the calibration file and applied together with the real-space probe step size.
\\
\textit{c. Find Bragg discs and centre the diffraction patterns}: a synthetic probe of matched disc radius is used as a correlation template to locate the Bragg discs at every scan position. The direct-beam (bright-field disc) position is measured at each pixel and a smooth function is fitted to it to correct the scan-dependent de-scan, after which the elliptical calibration is applied to correct for residual distortion in the diffraction plane.
\\
\textit{d. Perform the polar transform and compute radial statistics}: each calibrated pattern is transformed into polar (intensity versus scattering-angle) coordinates, and the local radial mean and variance are computed. The radial variance is retained as a \textit{hyperspy} signal because it is sensitive to the crystalline reflections while compressing each 4D dataset to a compact one-dimensional-per-pixel representation, which is well suited to pooling the signal across a given particle and across the ensemble of datasets. This is similar to the approach adopted in \cite{jinseok2026}.

For the collective analysis, each dataset is segmented into individual particles from its (virtual) annular dark-field image using a difference-of-Gaussians filter followed by Otsu thresholding, and per-particle morphology descriptors -- area, perimeter, eccentricity, aspect ratio, and major/minor axis lengths -- are extracted using the region-analysis routines of \textit{scikit-image}~\cite{scikit-image}. Particles touching the frame edge are excluded from the statistical analysis regarding particle shape. The radial-variance signal of each dataset is masked to only include particles and summed across the whole run to build the ensemble-averaged diffraction profile. With the edge cases included, we analysed 293 particles in total, reducing to 163 particles when edge-touching particles are excluded.

\begin{figure}
    \centering
    \includegraphics[width=\columnwidth]{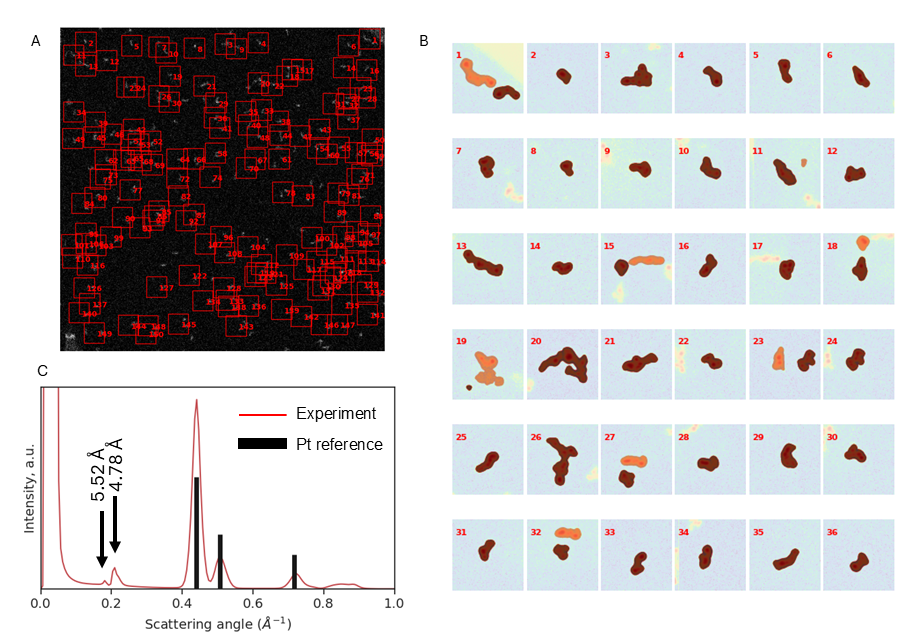}
    \caption{NBED dataset overview and diffraction analysis. \textbf{A} Survey HAADF-STEM image with all 153 numbered 4D-STEM acquisition regions overlaid as red boxes. \textbf{B} Gallery of segmentation masks for the first 36 nanoparticles, showing ADF signal (blue-green) and the automated particle mask (overlaid in red). \textbf{C} Azimuthal variance 1D diffraction profile summed over all valid datasets (red), overlaid with reference Pt peak positions (black bars). Arrows mark additional peaks at 5.52~\AA\ and 4.78~\AA\ d-spacings, attributed to an oxide phase.}
    \label{fig:fig_nbed1}
\end{figure}

The full set of 153 datasets collected from the selected region is shown in Figure~\ref{fig:fig_nbed1}\textbf{A}, with each numbered acquisition box overlaid on the survey image. Individual particle segmentation masks for all 153 acquisitions are provided in Supplementary Figure~\ref{fig:si_masks}. Panel \textbf{B} shows a gallery of the first 36 nanoparticles with their segmentation masks, illustrating the range of particle sizes and shapes captured. Panel \textbf{C} shows the azimuthal variance 1D diffraction profile summed across all valid datasets. The experimental curve (red) is compared with the reference peak positions for Pt (black bars). Two additional peaks are indicated by arrows at scattering angles corresponding to d-spacings of 5.52~\AA\ and 4.78~\AA, suggesting the presence of a surface oxide phase \cite{SALMERON1981207}. A complete gallery of the 1D diffraction profiles for all particles is shown in Supplementary Figure~\ref{fig:si_nbed2}, and the corresponding spatial maps of the dominant scattering vector for each dataset are given in Supplementary Figure~\ref{fig:si_nbed3}.

Crystallographic orientations were determined by automated crystal orientation mapping (ACOM). For each scan position the experimental Bragg-disc pattern was matched against a library of simulated face-centred-cubic Pt ($a = 3.92$~\AA) diffraction patterns generated with the diffraction-simulation and orientation-matching tools of \textit{Py4DSTEM}~\cite{py4dstem_paper}, and the best-fit orientation was stored as an orientation map. These maps were then analysed with the \textit{orix} library~\cite{orix} to produce inverse pole figures and pole density functions. Within each map, valid (on-particle) pixels were selected automatically using the product of the ACOM image-quality and confidence-index metrics, thresholded with Otsu's method, so that the orientation statistics are not contaminated by vacuum or weakly-diffracting pixels.

\begin{figure}
    \centering
    \includegraphics[width=\columnwidth]{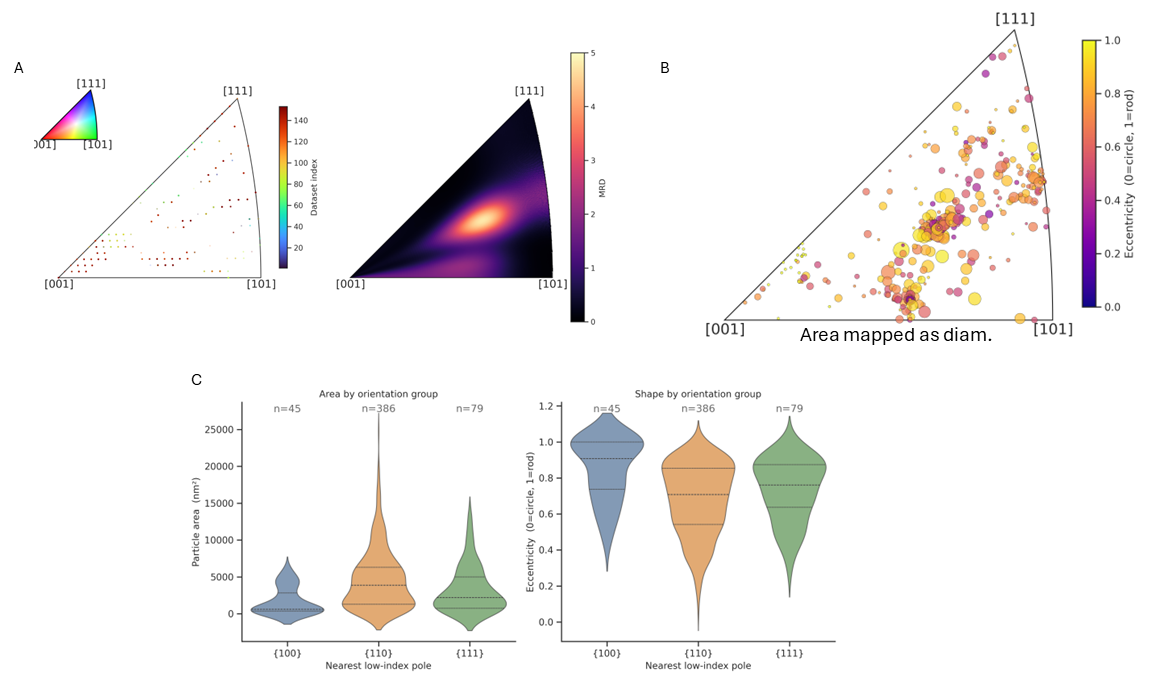}
    \caption{Automated crystal orientation mapping (ACOM) and morphology analysis from 153 NBED datasets. \textbf{A} Inverse pole figure (IPF) scatter plot of all particle orientations coloured by dataset acquisition index (left) and the corresponding pooled pole density function (right), revealing a weak \{110\} texture. \textbf{B} Per-particle IPF bubble chart: position encodes mean crystallographic orientation, marker size is proportional to particle area, and colour encodes eccentricity (0 = circular, 1 = rod-like). \textbf{C} Violin plots of particle area (left) and eccentricity (right) grouped by the nearest low-index pole (\{100\}, \{110\}, \{111\}); violin width is proportional to grain count in each group ($n$ shown above each violin). Note that some particles contained multiple crystallographic grains so the total number of grains analysed exceeds the number of particles detected.}
    \label{fig:fig_nbed2}
\end{figure}

The large number of datasets enables statistical analysis linking crystallographic orientation to particle morphology. Figure~\ref{fig:fig_nbed2} summarises the ACOM results, combined with the per-particle morphology descriptors introduced above. Panel \textbf{A} shows the inverse pole figure (IPF) scatter plot for all datasets coloured by acquisition order alongside the pooled pole density function, indicating a weak \{110\} texture (the peak is close to \{123\}). Panel \textbf{B} plots each particle's mean orientation in the IPF triangle with marker size proportional to particle area and colour encoding eccentricity, revealing that larger particles display a broader range of shapes. Panel \textbf{C} presents violin plots comparing particle area and eccentricity across the three nearest low-index pole groups (\{100\}, \{110\}, \{111\}); violin widths are scaled by particle count per group ($n$ indicated). Additional morphology scatter plots (area vs.\ aspect ratio and eccentricity) are provided in Supplementary Figure~\ref{fig:si_nbed1}. The texture analysis can be biased to the cases where the pattern quality was sufficiently high and more than two non-parallel Bragg vectors are present. However, similar observation can be made by examining the summed mean 1D patterns (see Figure~\ref{fig:si_mean_texture} where we see stronger \{111\} peak intensity than an ideal Pt powder pattern) or the strongest features in the 1D patterns across datasets (Figure~\ref{fig:si_nbed3}).

It is worth emphasising here how this method moves beyond standard morphology measurements such as particle size and shape to provide actionable insights into the surface chemistry of catalyst nanoparticles. For fuel cell applications, surface binding energies during the Oxygen Reduction Reaction (ORR) are strongly dictated by particle orientation and facet exposure details that conventional morphological analyses easily miss. When catalysts are intentionally designed to maximise specific active facets, such as the \{111\} surface, this method offers a direct way to verify whether that target surface coverage has actually been achieved \cite{Shi2021}. Furthermore, the technique could be applied to examine metal–support interactions, particularly for systems like platinum supported on ceramic oxides. These catalysts are known to achieve up to a threefold performance enhancement over standard commercial Pt/carbon black catalysts \cite{orr_PtCo_TEM}. Because the crystallographic alignment between the support and the nanoparticle is critical for driving specific reaction pathways, being able to characterise and optimise this relative orientation offers a significant advantage for catalyst engineering.

\subsection{Ptychography data collection and analysis}
\label{sec:analysis_ptycho}

The same automated data collection framework was also applied to collect ptychographic 4D-STEM data from the Pt nanoparticles. In this case the microscope was operated at 300~kV in a probe aberrations corrected configuration with a probe convergence semi-angle of approximately 26.7~mrad (defined by a 30~\textmu m condenser aperture), and the 4D-STEM data were again recorded on the MerlinEM detector. A real space step size of 1.95~\AA, dwell time of 1~ms, and probe defocus value of 40~nm were used to ensure sufficient probe overlap in successive probe positions. Pixel size of the detector used (MerlinEM quad) is 55~\textmu m , rotation angle between the scan and diffraction plane was -85.5 degrees, and the acceleration voltage 300~KeV. A total of 117 datasets were collected overnight, with both drift and focus correction applied after each acquisition. The survey image was collected at 250~kX, with the sampling of each 4D-STEM dataset set to be equivalent to 1.95~\AA\ step size. A comparison of a conventional HAADF-STEM image and a multi-slice ptychographic phase reconstruction from the same region is shown in Supplementary Figure~\ref{fig:si_stem_ptycho}, demonstrating the significantly improved signal-to-noise and phase sensitivity of the ptychographic reconstruction. The GUI workflow used for the ptychography data collection is shown in Supplementary Figure~\ref{fig:si_gui_ptycho}.

The ptychographic datasets were processed through two complementary reconstruction routes, both submitted as SLURM array jobs on the HPC cluster in the same templated fashion as the NBED analysis. First, a tilt-corrected bright-field (``parallax'') reconstruction was performed for every dataset using \textit{Py4DSTEM}~\cite{py4dstem_paper}. In addition to a rapid, robust real-space image, this reconstruction fits the aberration function of the illumination: fitting the contrast-transfer function yields the residual defocus (and low-order aberrations) for each acquisition \cite{varnavides2024iterativephaseretrievalalgorithms}, providing a per-dataset measure of beam quality that can be tracked across the multi-hour automated session. The estimated defocus value for each dataset based on the above algorithm over around 7 hours that the data collection routine was running is shown in Figure~\ref{fig:si_parallax}. The target defocus value of 40 nm defocus was achieved for the majority of the datasets, deviating for the latter cases, most likely limited by the accumulated contamination of the area used for defocus correction. Second,  multi-slice ptychographic reconstructions were performed with the \textit{PtyREX} package~\cite{ptyrex}. The reconstruction was performed with 5 slices in the object each 5 nm thick and 4 modes in probe model. For visualisation, the phase of the reconstructed exit-wave across the multiple slices was combined into a single image and assembled into the galleries presented below. The reconstructed probe (the incoherent sum over the mixed-state modes) was extracted for every dataset. 

\begin{figure}
    \centering
    \includegraphics[width=\columnwidth]{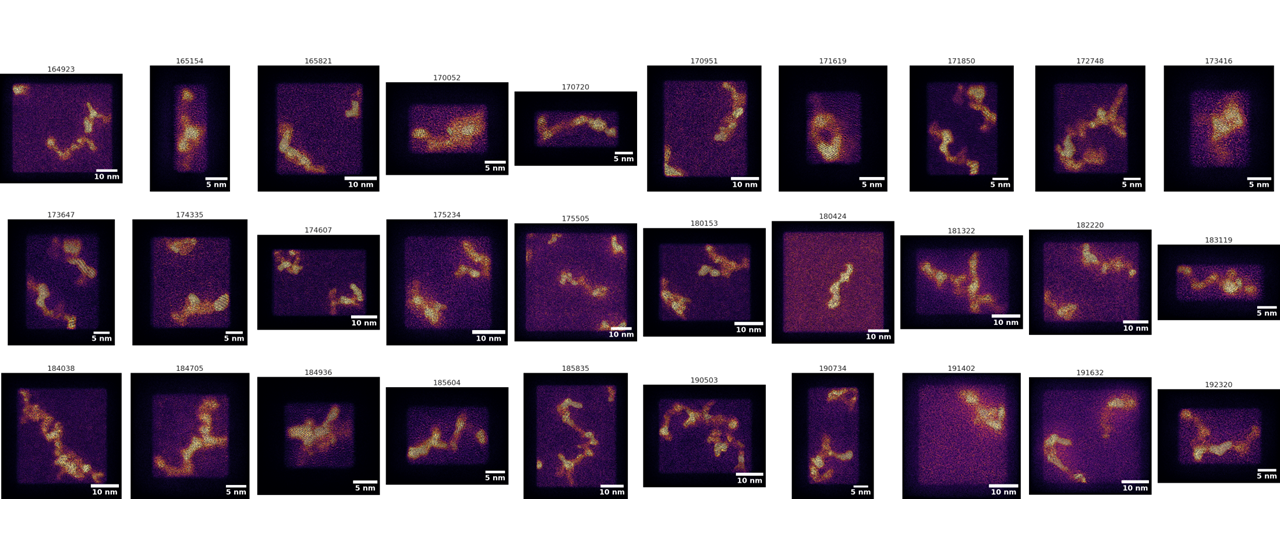}
    \caption{Ptychographic phase reconstructions of Pt nanoparticles from automated data collection, showing the first 30 datasets sorted by acquisition time. Acquisition timestamps are shown above each panel. Scale bars are shown in each panel. The variation in particle morphology, size, and number of visible nanoparticles per acquisition reflects the diversity of the sample region sampled by the automated routine.}
    \label{fig:fig_ptycho_grid}
\end{figure}

Figure~\ref{fig:fig_ptycho_grid} shows the ptychographic phase images for the first 30 reconstructed nanoparticles (sorted by acquisition time), with individual image contrast adjusted to maximise visibility. The full set of 117 reconstructions is provided in Supplementary Figure~\ref{fig:si_ptycho_recons}, and the corresponding reconstructed probe intensity images are shown in Supplementary Figure~\ref{fig:si_ptycho_probes}. Consistent with the parallax aberration fits, both the reconstructed object phase and the reconstructed probe show a progressive degradation over the course of the session. This is to be expected since no aberration correction -- apart from defocus -- was applied in the course of these acquisitions. The iterative ptychography algorithm can to some extent cope with these aberrations, but in extreme cases the reconstruction may only converge if probe is seeded with a prior. 

\begin{figure}
    \centering
    \includegraphics[width=\columnwidth]{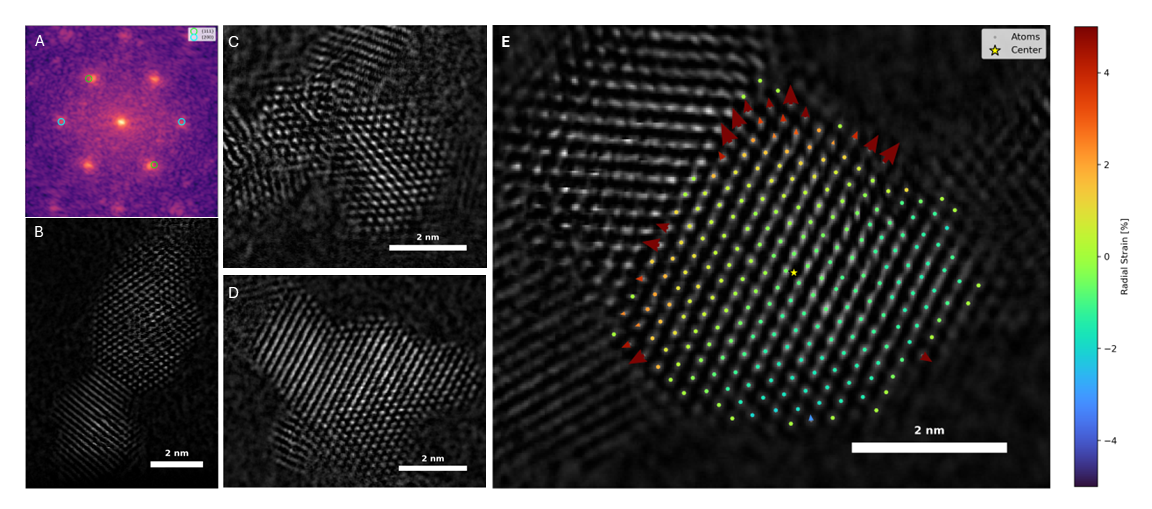}
    \caption{Atomic-resolution ptychographic analysis of Pt nanoparticles. \textbf{(A)} Fast Fourier transform (FFT) of the ptychographic reconstruction showing \{111\} and \{200\} Bragg reflections, confirming the [110] zone-axis orientation. \textbf{(B)--(D)} Multi-slice ptychographic phase reconstructions from three different nanoparticles, showing resolved atomic columns. Scale bars: 2~nm. \textbf{(E)} Atomic-resolution phase image a nanoparticle, with fitted atomic column positions coloured by the local radial lattice strain (\%). Red arrows indicate grain boundary regions.}
    \label{fig:fig_ptycho_analysis}
\end{figure}

To efficiently process the extensive volume of datasets generated during automated acquisition, a systematic computational pipeline was developed for the rapid identification and analysis of on-zone axis crystalline grains. The workflow employs a multi-step spatial-frequency filtering algorithm to isolate nanoparticles oriented along specific crystallographic zone axes. Generally, the algorithm analyses the Fast Fourier Transform (FFT) of the ptychography reconstruction phase images to identify a primary pair of characteristic reflections. An inverse Fast Fourier Transform (iFFT) masking procedure is then applied to selectively crop the specific regions exhibiting this periodicity. Within these isolated areas, a secondary FFT is performed to search for a subsequent pair of Bragg spots positioned at the correct symmetry angle relative to the first. As an example, Figure~\ref{fig:fig_ptycho_analysis}, Panel \textbf{A} shows the successfully filtered \{111\} and \{200\} reflections at the required angular orientation, confirming the [110] zone-axis orientation of the Pt nanoparticle. This high-throughput screening successfully extracts high-quality, on-zone ptychographic phase images of the Pt nanoparticles from the broader dataset (Figure~\ref{fig:fig_ptycho_analysis}, Panels \textbf{B--D}).
Following identification, the local structural deformations within the selected particles are mapped using an atom-tracking routine implemented in the Atomap library \cite{atomap}. By fitting the positions of individual atomic columns, the local lattice strain is calculated based on spatial deviations from the average interatomic spacing within the grain. Operating under the assumption that the nanoparticle undergoes predominantly radial deformation -- structural expansion or contraction directed relative to its centre -- these variations are then projected to generate high-resolution radial strain maps, Figure~\ref{fig:fig_ptycho_analysis}, Panel \textbf{E}, illustrating the strain field of one of the on-axis nanoparticles. This pipeline serves as an efficient screening mechanism to filter out high-quality datasets tailored to specific user requirements, enabling a rapid and preliminary assessment of the sample's structural characteristics to guide further investigation. 

This approach has valuable implications for atomic-scale strain analysis. As highlighted by \cite{pt_orr_strain}, lattice strain is fundamentally linked to oxygen binding energies through shifts in the d-band centre, which directly govern ORR activity. Performing high-precision strain mapping from STEM images manually is exceptionally laborious \cite{strain_NP_STEM}. The work proposed in this study represents one of the first automated acquisition routines capable of capturing the necessary structural data, laying the groundwork for fully automated strain analysis in future studies.

\section{Conclusions}
\label{sec:conclusions}
We have developed an automated data collection framework that applies a decision-making algorithm to identify the regions of a sample meeting a user-defined criterion, populates those regions with 4D-STEM acquisitions, at user-defined sampling conditions, and queues and collects them autonomously. The case demonstrated here identifies individual size-selectively deposited Pt nanoparticles dispersed on an amorphous carbon support by intensity thresholding and density-based clustering of a survey STEM image.

The framework was demonstrated in two complementary optical configurations. In nanobeam electron diffraction mode, 153 4D-STEM datasets were collected in a single unattended run. In focused-probe ptychography mode, 117 datasets were collected overnight and reconstructed to atomic resolution. In both cases the number of datasets and the fraction of beam time spent on regions of interest are well beyond what an operator could reasonably achieve manually, and the acquisitions are placed by a reproducible, recorded criterion rather than by operator judgement.

We have also presented a semi-automated analysis workflow that condenses this volume of data into interpretable microstructural information. Reducing each 4D NBED dataset to its azimuthal variance profile compresses the data by orders of magnitude while retaining sensitivity to the crystalline reflections, and makes it practical to pool the diffraction signal across the whole particle ensemble. Combining automated crystal orientation mapping with per-particle morphology descriptors allowed the orientation and shape distributions of the ensemble to be examined together -- an analysis that is only meaningful because of the number of particles sampled -- revealing a weak \{110\} texture across the population. For the ptychographic data, a high-throughput screening pipeline based on spatial-frequency filtering identified on-zone-axis particles automatically and enabled atomic-column-level strain mapping of the selected grains.

The results also make the current limitations explicit. Because no aberration correction beyond defocus was applied during the unattended runs, both the reconstructed probe and the object phase degrade measurably over the course of a multi-hour session. The reconstructions themselves provide a direct, quantitative record of this trend, which suggests an obvious next step: feeding the per-dataset aberration estimates back into the collection loop so that the illumination is periodically re-optimised rather than merely monitored.

Looking further ahead, the decision-making step is the most natural place to extend this work. In the present implementation, regions are selected from the intensity of a survey image alone, which is well matched to well-separated nanoparticles on a light support but carries no crystallographic information. We plan to base the selection criterion on the diffraction signal itself -- for example by acquiring a sparse or rapid 4D-STEM survey and triggering full-resolution acquisitions only where a particle is close to a chosen zone axis, or where a specific phase or degree of ordering is detected. A related target is the automated detection of grain boundaries, so that acquisitions can be placed preferentially on the interfaces between differently oriented grains rather than distributed uniformly over a particle. Such criteria would let the microscope seek out the comparatively rare configurations that are most scientifically informative, and would apply equally to more heterogeneous specimens than the one studied here. Because the hardware communication is isolated behind a common interface, this framework can be ported to other instruments, optical modes, and sampling conditions to expand from this work.

\section*{CRediT authorship contribution statement}

\noindent
\textbf{Mohsen Danaie:} Supervision, Conceptualization, Methodology, Software, Data curation, Formal analysis, Visualization, Validation, Resources, Writing -- original draft, Writing -- review \& editing.
\textbf{Max England:} Software, Methodology, Writing -- review \& editing.
\textbf{Yiming Xu:} Software, Methodology, Writing -- review \& editing.
\textbf{Ruomu Zhang:} Formal analysis, Visualization, Writing -- review \& editing.
\textbf{Ed Darnbrough:} Software, Methodology, Writing -- review \& editing.
\textbf{Josh Willem De Boer:} Formal analysis, Visualization, Writing -- review \& editing.
\textbf{Frederick Allars:} Formal analysis, Visualization, Writing -- review \& editing.
\textbf{Zaeem Najeeb:} Software, Methodology, Writing -- review \& editing.
\textbf{Aakash Varambhia:} Software, Methodology, Writing -- review \& editing.
\textbf{Jinseok Ryu:} Formal analysis, Visualization, Writing -- review \& editing.
\textbf{Benjamin Bradnick:} Software, Methodology, Writing -- review \& editing.
\textbf{Damien McGrouther:} Software, Methodology, Writing -- review \& editing.
\textbf{Manfred E. Schuster:} Software, Methodology, Writing -- review \& editing.
\textbf{Christopher S. Allen:} Supervision, Conceptualization, Methodology, Writing -- review \& editing.

\section*{Declaration of competing interest}
The authors declare that they have no known competing financial interests or personal relationships that could have appeared to influence the work reported in this paper.

\section*{Data availability}
The automated data collection software described in this work is openly available under an open-source licence at \url{https://github.com/ePSIC-DLS/Automated-Data-Collection} \cite{epsic}. The analysis notebooks used to produce the results and figures presented here are available at \url{https://github.com/M0hsend/Automated_STEM_GUI_Manuscript} \cite{epsic_analysis}. The raw 4D-STEM datasets were acquired at ePSIC, Diamond Light Source and are archived under the Diamond Light Source data policy.

\section*{Acknowledgements}
We thank Diamond Light Source for access and support in use of the electron Physical Science Imaging Centre (Instrument E02 and proposal number mg44468) that contributed to the results presented here. The Pt nanoparticle sample was kindly prepared for us by Alex Large and Henry Hoddinott at the cluster source facility at B07, Diamond Light Source, overseen by Georg Held and Richard Palmer.

\clearpage
 \bibliographystyle{elsarticle-num}
 \bibliography{cas-refs}

\clearpage
\appendix
\section{Note on calibrations}
\label{sec:appendix_cals}
In order to calibrate the real and reciprocal pixel sizes we used the standard evaporated gold cross-grating sample. We used the native JEOL scan controls for triggering the MerlinEM acquisitions in this case with reference 4D-STEM data collected at the 150~kx magnification, similar to the FOV of the survey image. Similar workflow as the one discussed in the \ref{sec:analysis} was applied to the calibration data collected from evaporated gold on carbon, with the difference that in the end the reciprocal pixel size was measured by fitting to the azimuthal integration of the gold ring pattern. Furthermore, an affine transform was calculated to correct for any deviations from the perfect diffraction roundness in the diffraction plane. After fitting we calibrated the reciprocal pixel size to 0.005654~\text{\AA$^{-1}$} and the affine transform to be applied for roundness correction to be [301.686, 319.484, 127.796, 122.917, -2.892]. The calibration notebook is included in the github repo of this manuscript \cite{epsic_analysis}.

\section{Metadata collected}
\label{sec:appendix_metadata}
\begin{figure}
    \centering
    \includegraphics[width=\columnwidth]{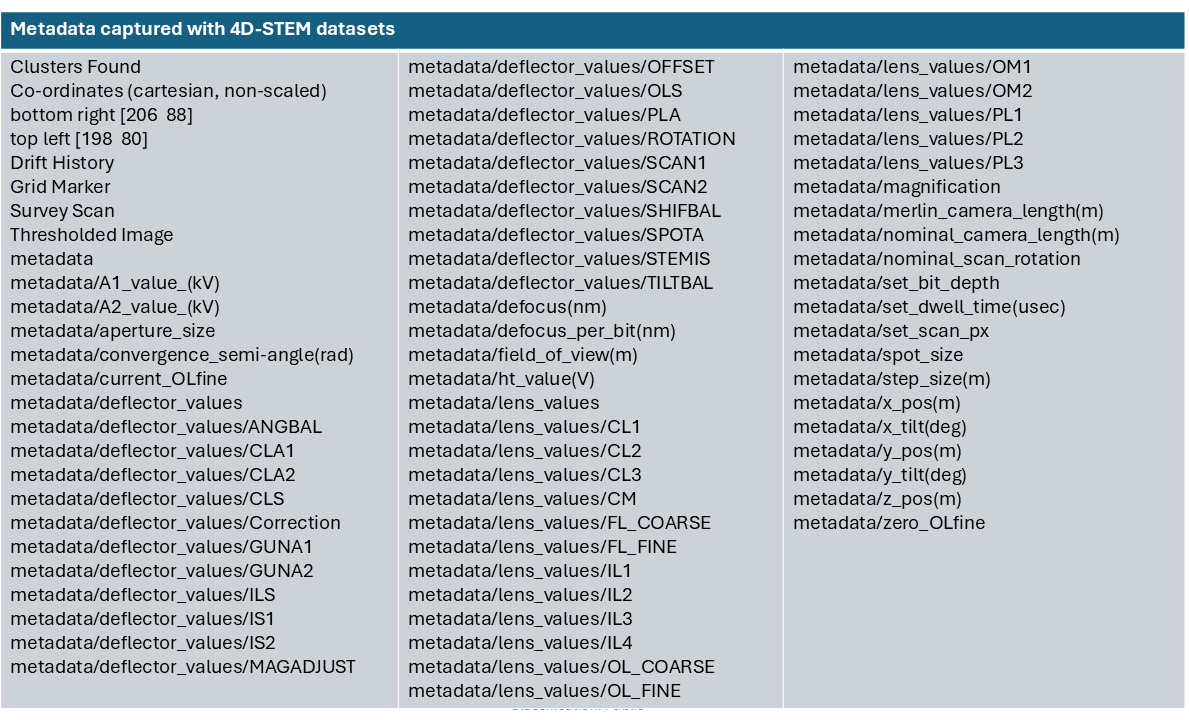}
    \caption{Complete list of metadata items captured with each automated 4D-STEM dataset, covering microscope state, scan parameters, detector settings, stage coordinates, and the images used for automated data collection decisions.}
    \label{fig:metadata}
\end{figure}

The table in Figure~\ref{fig:metadata} lists the items captured in the metadata of the 4D-STEM datasets collected. Metadata includes microscope state (lens excitations, deflector values, accelerating voltage, spot size, aperture), scan parameters (step size, field of view, dwell time, bit depth), stage coordinates, nominal and calibrated camera length, drift history, and the survey and thresholded images used for automated region selection.

\section{Supplementary figures}
\label{sec:appendix_si}

\begin{figure}
    \centering
    \includegraphics[width=\columnwidth]{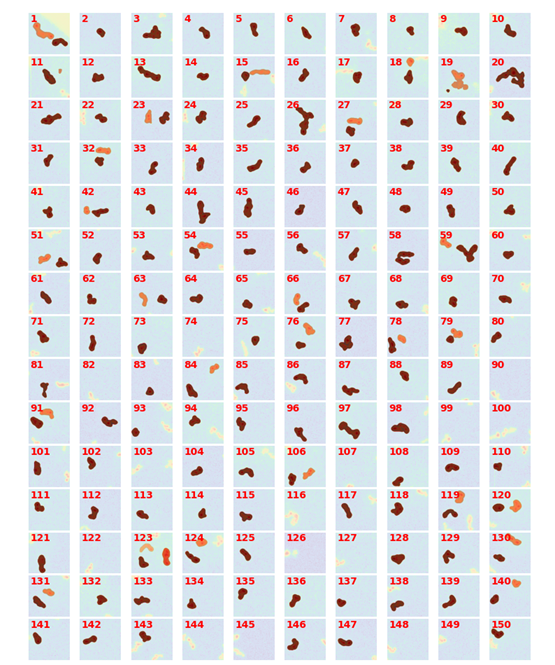}
    \caption{Segmentation masks for all 153 automatically acquired NBED datasets, numbered by acquisition order. Each panel shows the ADF signal (blue-green) overlaid with the automated particle mask, illustrating the range of nanoparticle sizes, shapes, and the occasional presence of multiple particles per acquisition.}
    \label{fig:si_masks}
\end{figure}

\begin{figure}
    \centering
    \includegraphics[width=\columnwidth]{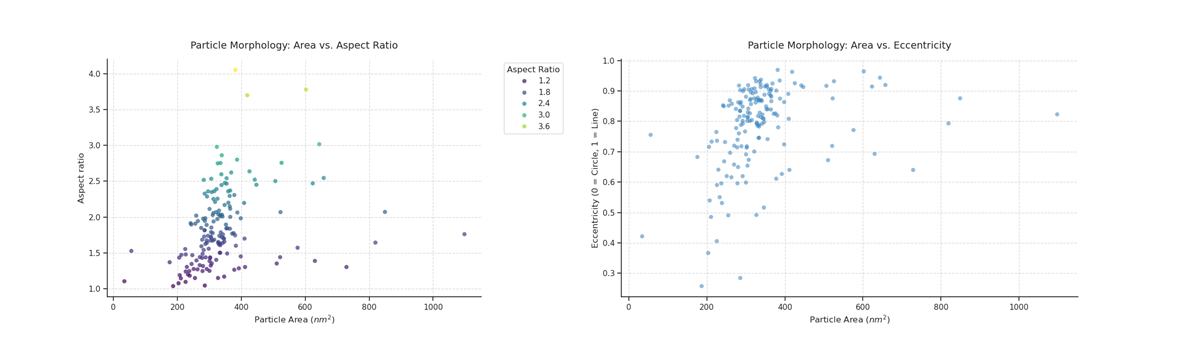}
    \caption{Particle morphology scatter plots from NBED analysis. \textbf{Left}: particle area (nm$^2$) vs.\ aspect ratio, with colour encoding aspect ratio magnitude. \textbf{Right}: particle area vs.\ eccentricity (0 = circular, 1 = rod-like). The majority of particles fall in the 100--500~nm$^2$ area range with aspect ratios between 1 and 2.5.}
    \label{fig:si_nbed1}
\end{figure}

\begin{figure}
    \centering
    \includegraphics[width=\columnwidth]{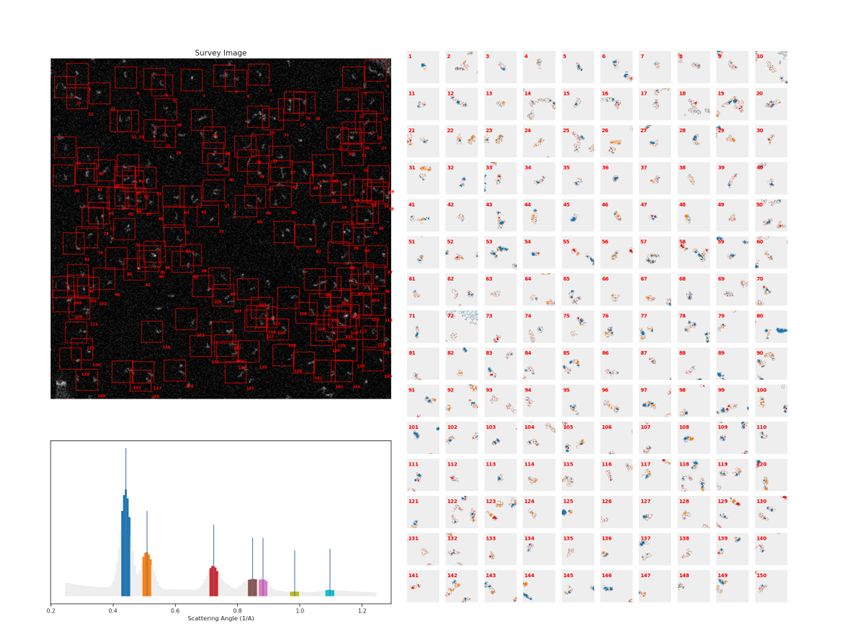}
    \caption{Overview of NBED 1D diffraction profiles for all 150 datasets. \textbf{Left}: survey image with numbered acquisition boxes and stacked bar chart of dominant scattering angle per dataset. \textbf{Right}: gallery of 1D azimuthal variance profiles for all 150 particles, coloured by peak scattering angle, enabling rapid identification of datasets with anomalous or oxide-phase diffraction signal.}
    \label{fig:si_nbed2}
\end{figure}

\begin{figure}
    \centering
    \includegraphics[width=\columnwidth]{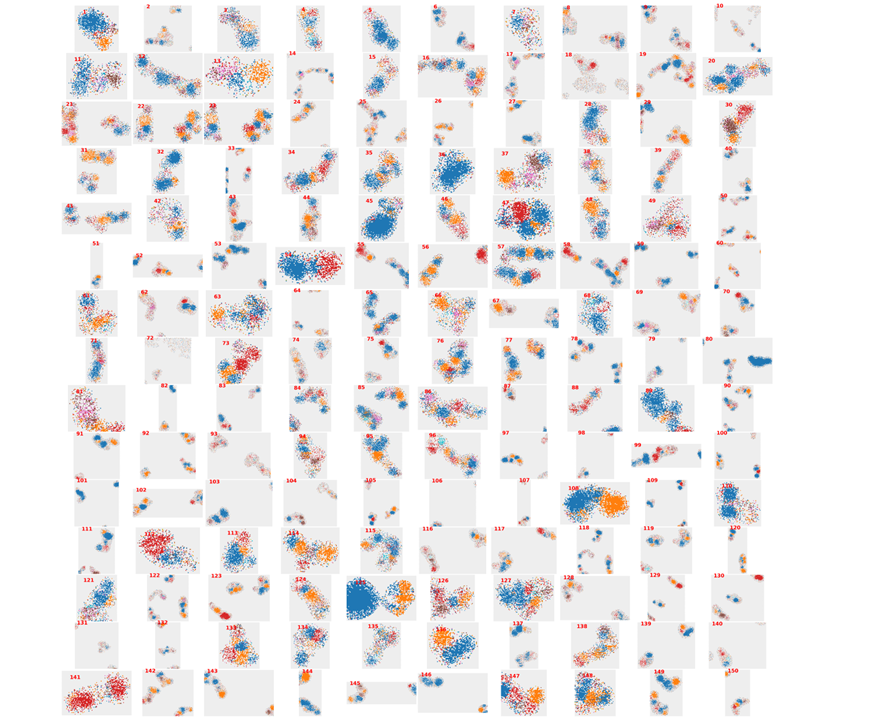}
    \caption{Spatial maps of the dominant scattering vector for all 150 NBED datasets. Each panel shows, for every pixel of the segmented particle, the scattering angle at which the azimuthal variance profile peaks, colour-coded as in Figure~\ref{fig:si_nbed2}. Datasets are ordered by acquisition sequence.}
    \label{fig:si_nbed3}
\end{figure}

\begin{figure}
    \centering
    \includegraphics[width=\columnwidth]{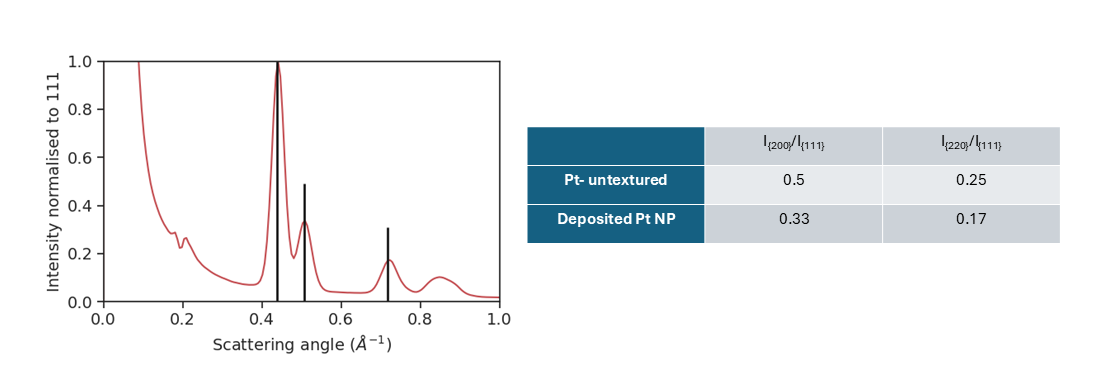}
    \caption{Summed mean diffraction signal from all the NP's, overlaid with ideal fcc Pt powder peak positions and intensities. Table on the right compares the observed peak intensities with the untextured pattern.}
    \label{fig:si_mean_texture}
\end{figure}

\begin{figure}
    \centering
    \includegraphics[width=\columnwidth]{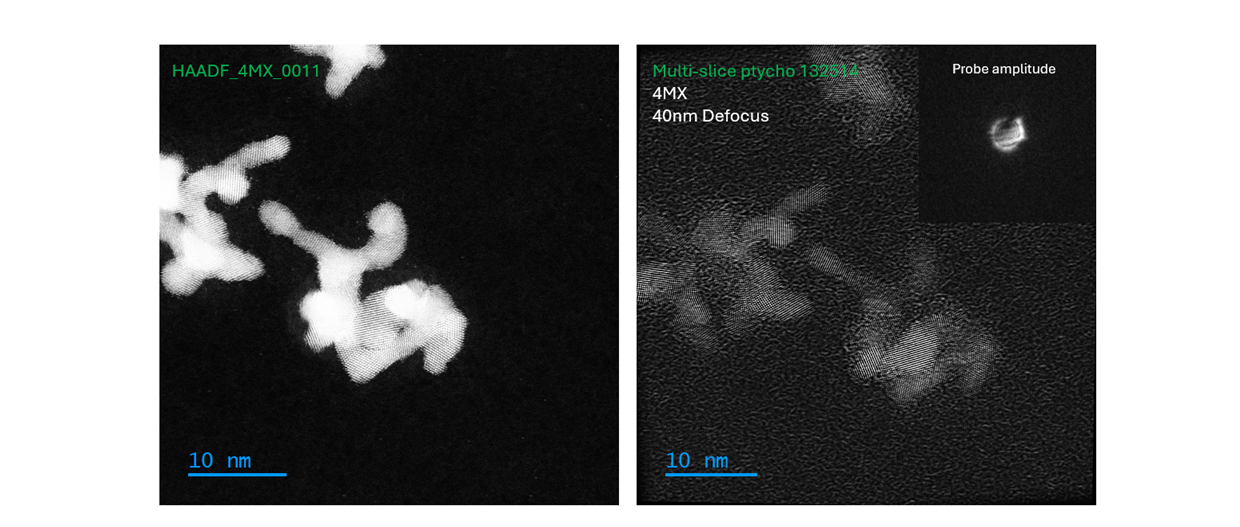}
    \caption{Comparison of conventional HAADF-STEM (\textbf{left}) and multi-slice ptychographic phase reconstruction (\textbf{right}) of the same Pt nanoparticle cluster. The ptychographic reconstruction (40~nm defocus, 4~MX camera length) reveals significantly improved phase contrast and signal-to-noise compared with the incoherent HAADF image. Inset in the right panel shows the reconstructed probe amplitude. Scale bars: 10~nm. The same sampling conditions were matched in the automated routine.}
    \label{fig:si_stem_ptycho}
\end{figure}

\begin{figure}
    \centering
    \includegraphics[width=\columnwidth]{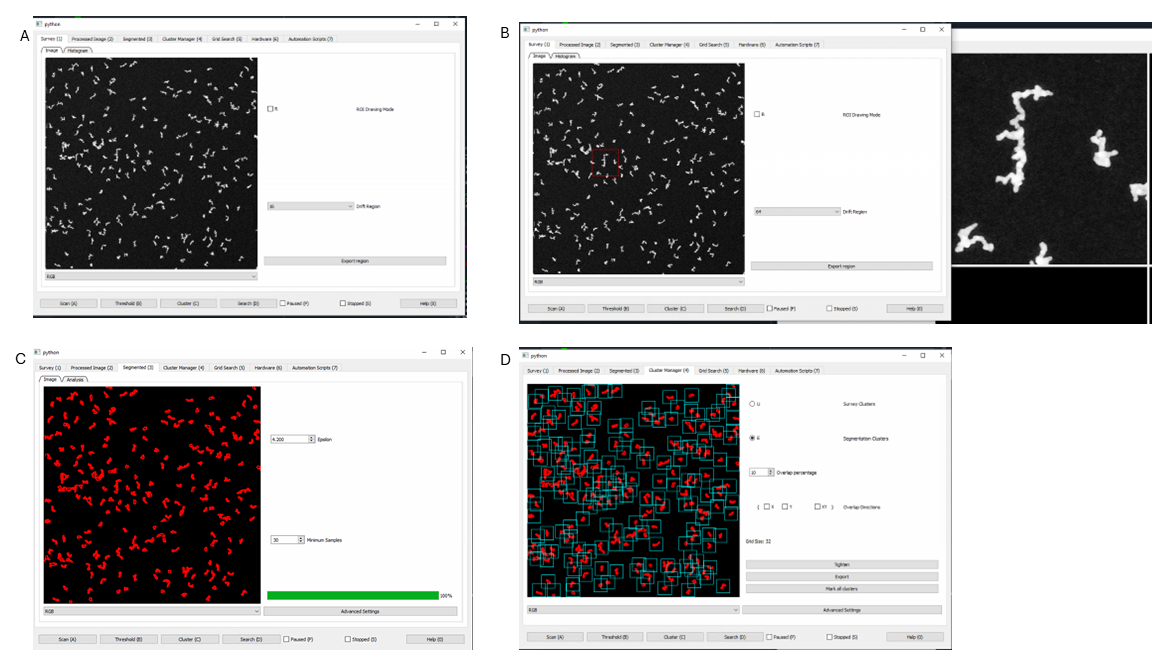}
    \caption{GUI screenshots for the ptychography automated data collection session. \textbf{A} Survey HAADF-STEM image of the Pt nanoparticle sample. \textbf{B} Thresholded image with a user-defined circular ROI (red dashed circle) restricting data collection to the central field of view; the cropped binary image is shown to the right. \textbf{C} DBSCAN-clustered image with individual particle clusters in red. \textbf{D} Populated 4D-STEM acquisition grid overlaid on the segmented regions. Region showed in the upper right is used for focus and drift correction.}
    \label{fig:si_gui_ptycho}
\end{figure}

\begin{figure}
    \centering
    \includegraphics[width=\columnwidth]{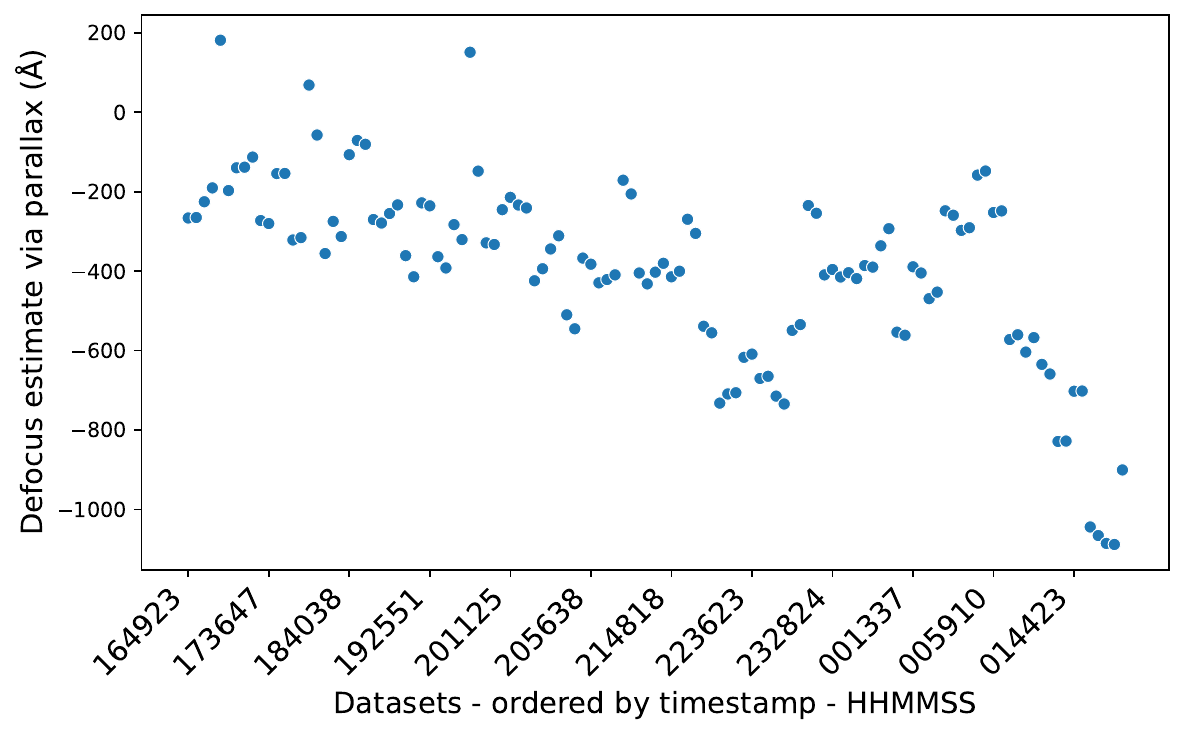}
    \caption{Estimated defocus of the probe for each dataset using the parallax algorithm. }
    \label{fig:si_parallax}
\end{figure}

\begin{figure}
    \centering
    \includegraphics[width=\columnwidth]{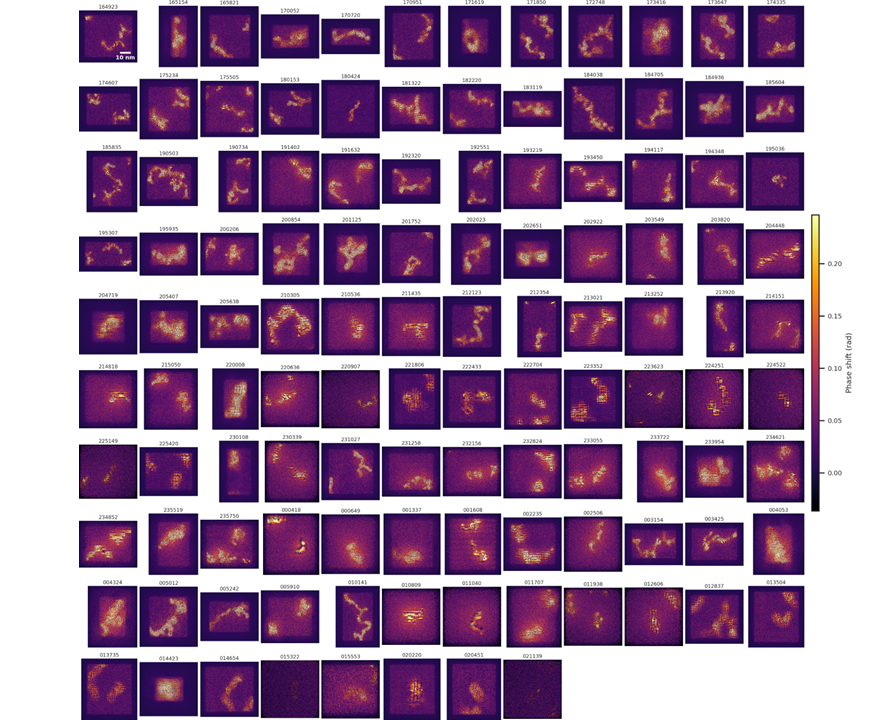}
    \caption{Ptychographic phase reconstructions for all automatically acquired datasets, sorted by acquisition time (earliest top-left). Timestamps are shown above each panel. A shared phase shift colour scale (rad) is shown on the right. Reconstructions collected later in the session show progressively reduced quality, consistent with gradual build-up of probe aberrations over the course of the multi-hour automated acquisition.}
    \label{fig:si_ptycho_recons}
\end{figure}

\begin{figure}
    \centering
    \includegraphics[width=\columnwidth]{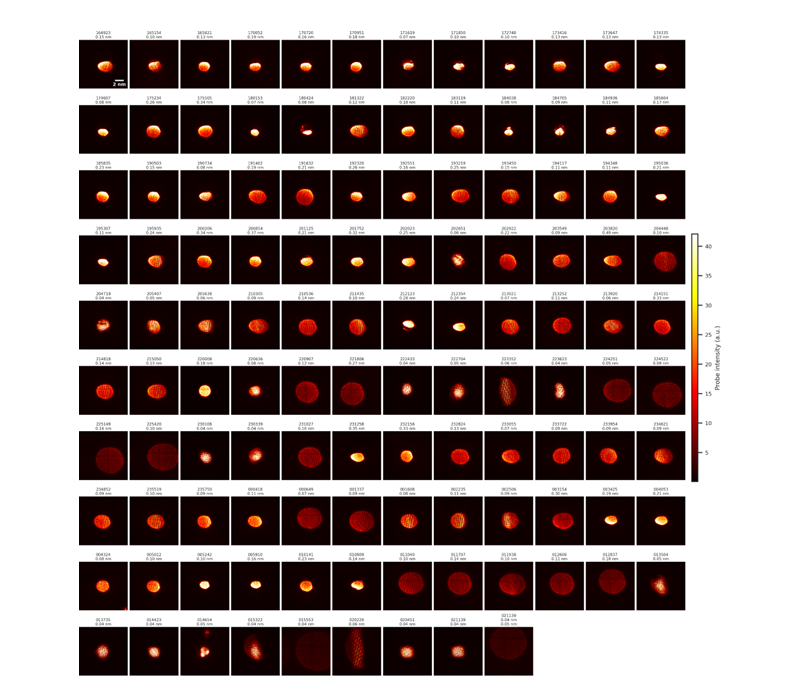}
    \caption{Reconstructed probe intensity ($|\text{probe}|^2$, incoherent sum of all modes) for all ptychographic datasets, sorted by acquisition time. The estimated probe FWHM (nm) is annotated below each timestamp. Probe intensity and size remain relatively stable during the early acquisitions and degrade progressively over time, providing a direct measure of beam quality evolution during the automated session. A shared colour scale (probe intensity, a.u.) is shown on the right.}
    \label{fig:si_ptycho_probes}
\end{figure}

%%\newpage
%% \bibliographystyle{elsarticle-num} 
%% \bibliography{cas-refs}





% \end{thebibliography}
\end{document}